\documentclass[%
 reprint, 
 superscriptaddress,
 amsmath,amssymb,
 aps, prx,
]{revtex4-2}

\usepackage{graphicx}
\usepackage{bm} 
\usepackage{upgreek} 
\usepackage{soul}

\usepackage{xcolor}

\graphicspath{{Figures/}}

\newcommand{\flux}[1]{\nu_{#1}}
\newcommand{\vmin}{{\alpha}}
\newcommand{\Res}{\mathop{\mathrm{Res}}}
\newcommand{\bbE}{\mathop{\mathbb{E}}}

\newcommand{\EqRef}[1]{Eq.~\eqref{#1}}
\newcommand{\FigRef}[1]{Fig.~\ref{#1}}

\definecolor{lucacolor}{rgb}{0.5,0.8,0.3}

\begin{document}


\title{Non-stationary dynamics of interspike intervals in neuronal populations}

\author{Luca Falorsi}
\email{luca.falorsi@gmail.com}
\affiliation{PhD program in Mathematics, Sapienza Univ. of Rome, Piazzale Aldo Moro 5, Rome, Italy}
\affiliation{Natl. Center for Radiation Protection and Computational Physics, Istituto Superiore di Sanit\`a, Viale Regina Elena 299, Rome, Italy}

\author{Gianni V. Vinci}
\affiliation{Natl. Center for Radiation Protection and Computational Physics, Istituto Superiore di Sanit\`a, Viale Regina Elena 299, Rome, Italy}

\author{Maurizio Mattia}
\email{maurizio.mattia@iss.it}
\affiliation{Natl. Center for Radiation Protection and Computational Physics, Istituto Superiore di Sanit\`a, Viale Regina Elena 299, Rome, Italy}

\date{\today}

\begin{abstract}
We study the joint dynamics of membrane potential and time since the last spike in a population of integrate-and-fire neurons using a population density framework. 
This leads to a two-dimensional Fokker–Planck equation that captures the evolution of the full neuronal state, along with a one-dimensional hierarchy of equations for the moments of the inter-spike interval (ISI). 
The formalism allows us to characterize the time-dependent ISI distribution, even when the population is far from stationarity, such as under time-varying external input or during network oscillations. 
By performing a perturbative expansion around the stationary state, we also derive an analytic expression for the linear response of the ISI distribution to weak input modulations.
\end{abstract}

\maketitle

\section{Introduction}

Neurons communicate through stereotyped, spike-like electrical impulses generated when their membrane potentials exceed a threshold. 
The intervals between consecutive spikes known as interspike intervals (ISIs), are highly irregular in the cortex, even under identical stimulus/environmental conditions \cite{Softky1993, Shadlen1994}. 
Such variability gives rise to rich ISI statistics and non-Poisson distributions \cite{Bair1994, Maimon2009, Swindale2023}, suggesting that information transmission by cortical neurons may be fundamentally constrained by noise \cite{Shadlen1998}. 
Nevertheless, stimulus-dependent modulations of ISI statistics can be faithfully represented at the population level, enabling temporally precise encoding of information \cite{Bair1996, Shadlen1998}. 
From this perspective, developing a theoretical framework that explains how ISI statistics unfold in neuronal populations is crucial for understanding the principles of neural information processing and encoding in the brain.

Variability in neuronal spike patterns arises from the incessant fluctuations of input currents driven by both continuous changes in the sensed environment and intrinsic brain activity, even under constant external stimulation \cite{Arieli1996, Faisal2008}. 
These fluctuations represents an unavoidable source of noise that can be further amplified by neurons acting as excitable systems \cite{Engel2008}, where stochastic resonance may shape and enrich the ISI distribution \cite{Longtin1991}. 
During sleep or general anesthesia, when the unconscious brain enters an isolated and synchronized state, slow waves generate a quasi-periodic alternation between high-firing and silent phases \cite{SanchezVives2017}, leading to distinctive ISI distributions \cite{Vyazovskiy2009} that differ from those observed during wakefulness across cortical layers and regions \cite{DeKock2008, Shinomoto2009}. 
The diversity of ISI statistics has been theoretically linked to the presence of metastable dynamics in neuronal systems, both during spontaneous activity \cite{Wilbur1983, LitwinKumar2012} and during performance of specific cognitive tasks \cite{Zipser1993, Compte2003}. 
Therefore, a comprehensive theory of ISI dynamics must account for neuronal populations operating far from equilibrium.

The study of ISI distributions in stationary neuronal populations has long been a cornerstone of stochastic analysis in neuroscience \cite{Gerstein1964, Ricciardi1979, Tuckwell1988}. 
At the single-neuron level, ISIs naturally arise within the framework of the \emph{first-passage time} (FPT) problem, which characterizes the distribution of times required for the membrane potential $V(t)$ to reach threshold. 
Under the diffusion approximation, the subthreshold dynamics of an integrate-and-fire (IF) neuron model is governed by the stochastic differential equation
\begin{align}
    \tau_m\,dV = f(V)\,dt + \mu\,dt + \sqrt{2\tau_m D}\,dW,
\end{align}
where $f(V)$ is the intrinsic drift, $\mu(V,t)$ and $\sigma(V,t)^2 = 2 D(V,t)$ denote the infinitesimal mean and variance of the synaptic input, and $W$ is a Wiener process with 0 mean and $\langle dW(t) dW(t')\rangle=\delta(t-t') dt$. 
Setting $\tau_m=1$ as unit measure of time, the corresponding membrane potential density $p^\text{A}(v,t)$ satisfies the Fokker-Planck (FP) equation
\begin{align}
    \partial_t p^\text{A}(v,t) &= \mathcal{L} \, p^\text{A}(v,t)\end{align}
where $\mathcal{L} \, p^\text{A} = -\partial_v[(f+\mu)p^\text{A} - D\partial_v p^\text{A}]$ is the FP operator, and the emission of a spike due to the reaching of the threshold $\theta$ is modelled by an absorbing barrier ($p^\text{A}(\theta,t) = 0$).
For the sake of generality, the membrane potential is lower-bounded including a reflecting barrier in $\alpha$ where the flux defined as $- \partial_v S = \mathcal{L} \, p^\text{A}$ is set to $S(\alpha,t) = 0$.
The superscript ``A'' indicates the absorbing boundary condition without the reinjection term we will introduce later.

In the classical formulation, the ISI distribution is obtained by solving the above FP equation with initial condition $V(0) = y$:
\begin{align}\label{time-isi-single}
   \partial_t p^\text{A}(v,t) &= \mathcal{L} \, p^\text{A}(v,t), \\
   p^\text{A}(v,0) &= \delta(v-y),
\end{align}
The FPT (or ISI) density is then the flux of realization crossing the threshold emitting a spike
\begin{align}
    \rho(t|y) = \left.-D(v,t) \, \partial_v p^\text{A}(v,t) \right|_{v=\theta} =  S(\theta,t) \, .
\end{align}
For stationary currents ($\dot{\mu} = \dot{D} = 0$), exact ISI probability densities exist only for a few simplified models \cite{Gerstein1964, Ricciardi1988, Tuckwell1988}. 
However, Siegert's recursion formula \cite{Siegert1951, Tuckwell1988} has enabled derivation of the first two moments of the ISI for a broader class of IF neurons \cite{Ricciardi1979, Fusi1999, Brunel2000, Lindner2003}, even in the presence of multiplicative noise, as in conductance-based models \cite{Musila1994, Sanzeni2022}. 
These results have been partially recovered through spectral decomposition of the Fokker-Planck operator \cite{Ricciardi1988, Pietras2020, Vinci2024}. 
Collectively, they indicate a degree of universality in ISI distributions across IF models \cite{Vilela2009, Ostojic2011}.

Although this framework is well established for stationary inputs, its extension to \emph{non-stationary} or \emph{self-consistent} regimes introduces substantial difficulties. 
In mean-field descriptions of spiking neuron networks \cite{Amit1997}, the mean and variance of synaptic currents depend on the population firing rate---which is intimately tied to ISI statistics---rendering the evolution operator $\mathcal{L}$ both time-dependent and nonlinear \cite{Brunel1999, Mattia2002}.
Self-consistency also necessitates modeling the reset of the membrane potential after spike emission. 
This is achieved by reinjecting the outgoing flux $\nu(t) = S(\theta,t)$---the population firing rate---at the reset potential $H$ as an additional source term:
\begin{equation}
   \partial_t p(v,t) = \mathcal{L} \, p(v,t) + \nu(t)\,\delta(v-H) \, .
\end{equation}
In this \emph{population density approach} (PDA) \cite{Knight1996, Brunel1999, Nykamp2000, Knight2000, Mattia2002}, individual realizations are preserved, and unlike $p^\text{A}(v,t)$, the probability density $p(v,t)$ remains normalized at all times ($\int_\alpha^\theta p(v,t), dv = 1$). 
This continuous coupling of subthreshold voltage dynamics to spike-time history transcends the renewal approximation. However, the approach comes at the cost of severing any direct link to single-neuron ISI statistics.

Alternative approaches linking population dynamics to single-neuron ISIs include the \emph{spike response model} (SRM) formalism \cite{Gerstner1995} and the \emph{refractory density method} (RDM) \cite{Chizhov2006}. 
The SRM formulation computes population activity by filtering past firing rates with a state-dependent ISI distribution. 
The RDM, in contrast, explicitly evolves the population density structured by the age $\tau$ (time since last spike), employing a hazard function derived from the stationary ISI distribution. 
To address non-stationarity, both frameworks break the strict renewal hypothesis by modulating these renewal quantities based on the instantaneous input statistics $\mu(t)$ and $D(t)$. 
However, this relies on a \emph{quasi-renewal} approximation: it assumes that at any instant, the spiking statistics match those of a stationary state driven by the current inputs. 
Far from steady state, this approximation fails because the true hazard rate and ISI statistics depend nontrivially on the full voltage dynamics, which are shaped by the history of time-varying, self-consistent inputs. 
Consequently, the actual non-stationary quantities differ from the stationary renewal ones, and, as we will show, it is impossible to fully decouple the age-dependent dynamics from the underlying voltage evolution.

To address all these issues, here we formulate the neuronal dynamics in an extended state space that jointly tracks the membrane potential $V$ and the time since last spike $\tau$. 
The coupled stochastic dynamics reads
\begin{align}\label{eq:coupling}
\begin{cases}
    dV = f(V)\,dt + \mu\,dt + \sqrt{2 D} \, dW \\
    d\tau = dt
\end{cases} \, ,
\end{align}
with reset conditions $V(t^-) = \theta \Rightarrow V(t)=H$ and $\tau(t)=0$. 
This construction leads naturally to a two-dimensional FP equation for the joint density $q(v,\tau,t)$, enabling the analysis of ISI statistics in non-stationary, self-consistent regimes beyond the reach of classical renewal formulations and their approximated extensions.

\section{Density equation}

At the population level, the joint density $q(v,\tau,t)$ evolves according to the two-dimensional PDE
\begin{align}
\label{eq:isi-fp}
\partial_t q(v,\tau,t)
    = & \mathcal{L} \, q(v,\tau,t) \notag
       - \partial_\tau q(v,\tau,t)\\
       &+ \delta_H(v)\,\delta_0(\tau)\,\nu(t),
\end{align}
with absorbing boundary condition $q(\theta,\tau,t) = 0$ for all $t$ and $\tau\ge 0$.  
Here $\mathcal{L}$ acts on the voltage variable $v$.  
We define the partial flux through threshold at ISI $\tau$ as

\footnote{In general, given a function $f(v, \cdot)$ vanishing at the boundary $v=\theta$, we define $\flux{f}(\cdot) := - D(t)\partial_{v} f(v, \cdot)|_{v=\theta}$ as the partial flux at the boundary.}
:
\begin{align}
    \flux{q}(\tau,t)
        := -D(t)\,\partial_v q(v,\tau,t)\big|_{v=\theta},
\end{align}
so that the total firing rate is given by
\begin{align}\label{eq:nuq}
    \nu(t)
        = \int_{0}^{\infty} \flux{q}(\tau,t)\,d\tau.
\end{align}
The ISI distribution at time $t$ is therefore
\begin{equation}
    \rho(\tau,t) = \frac{\flux{q}(\tau,t)}{\nu(t)}.
\end{equation}

\subsection{Relationship with age-structured population dynamics}

Our approach clarifies the relationship between the population density method and age-structured population dynamics, providing a systematic framework for assessing the validity of the quasi-renewal approximation used in these models.
We can establish a connection between the introduced formalism and the SRM \cite{Gerstner1995}. 
Consider the initial distribution $q(v,\tau,t=0) = \delta_{H}(v)\delta_0(\tau)$ (we refer to the Appendix for the general case). 
If we then rewrite $q(v,\tau, t) = g(v, \tau, t)[\nu(t - \tau)], \forall \tau < t$ and substitute this ansatz into equation \eqref{eq:isi-fp} we find that $g(v, \tau, t)$ satisfied the partial differential equation:
\begin{align}\label{eq:g-fp}
\partial_t g(v,\tau,t) =& \mathcal{L} \, g(v,\tau,t) - \partial_\tau g(v,\tau,t) + \\ & +\delta_H(v)\delta_0(\tau) \quad \forall \tau <t \notag.
\end{align}
Notice that $g$ depends on the past firing rate only through the mean $\mu(t)$ and the variance $D(t)/2$ of the synaptic current.
The function $g(v,\tau,t)$ describes the probability density of a neuron that fired at time $t-\tau \ge 0$. We can then express the firing rate $\nu(t)$ as a function of the past activity: 
\begin{align} \nu(t) = \int_0^t\flux{g}(\tau, t)\nu(t-\tau) d\tau + \rho(t|H). 
\end{align}
The last term represents the fraction of neurons firing for the first time, with an ISI of $t$, and progressively becomes irrelevant at $t$ becomes larger. 
In the SRM framework, the integral kernel $\nu_g(\cdot, t)$ is typically approximated using the stationary ISI distribution. In contrast, our formulation provides an exact expression, at the cost of solving a two-dimensional equation.

Similarly, marginalizing the population density equation \eqref{eq:isi-fp} over the membrane potential $v$ yields the RDM formulation \cite{Schwalger2019}:
\begin{align}\label{eq:refractory}
\partial_t r(\tau, t) = - \partial_\tau r(\tau, t) - \nu_q(\tau, t) + \nu(t)\delta_{0}(\tau),
\end{align}
where $
r(\tau, t) := \int_{\vmin}^{\theta} q(v, \tau, t), dv.
$
This marginal equation, however, remains unclosed, as the RDM approximates the flux term $\nu_q$ through a state-dependent stationary hazard function.


\subsection{Moment equations}
The joint density equation \eqref{eq:isi-fp} can be simplified by taking the Laplace transform on the variable $\tau$. The time evolution of the Laplace transform $\tilde{q}(v, s, t) = \int_{0^-}^\infty e^{-s\tau} q(v,\tau, t) d\tau,\ s\in \mathbb{C}$ is then given by the equation: 
\begin{align}\label{isi-fp-laplace}
    \partial_t \tilde{q}(v, s, t) = \mathcal{L} \, \tilde{q}(v, s, t) - s\tilde{q}(v,s,t) + \delta_H(v)\nu(t). 
\end{align}
A key feature of \eqref{isi-fp-laplace} is that, for each fixed Laplace frequency $s$, the evolution of $\tilde{q}(v,s,t)$ is decoupled from that at other values $s' \neq s$; the only external drive enters through the population firing rate $\nu(t)$.
By computing the Taylor expansion of equation \eqref{isi-fp-laplace} we can further derive a hierarchy of one-dimensional PDEs for the moments $T_n(v,t):=  \int_{0^-}^\infty \tau^n q(v,\tau,t)d\tau$:
\begin{align}\label{eq:moments}
    \partial_t T_0(v,t) &= \mathcal{L} \, T_0(v,t) + \delta_H(v)\nu(t) \\
    \partial_t T_n(v,t) &= \mathcal{L} \, T_n(v,t) + nT_{n-1}(v,t) \quad \forall n \ge 1
\end{align}
with absorbing boundary at $\theta$. Then the flux of $T_n(\cdot, t)$ at the threshold gives the $n$-th moment of the ISI of neurons firing at time $t$:
\begin{align}
    m_n(t):=\!\!\!\bbE_{\tau\sim\rho(\tau,t)}\!\!\!\!\!\![\,\tau^n] = \dfrac{-D(t)}{\ \nu(t)}\partial_vT_n(v,t)|_{v=\theta} = \dfrac{\flux{T_n}(t)}{\nu(t)}
\end{align}
This hierarchy of equations is one of the main results of this work, as it allows studying the evolution of moments of the ISI distributions when the population is far from the stationary state. 
\section{Stationary state}
We begin by analyzing the stationary solution $q^0(v,\tau)$ of the ISI population density equation~\eqref{eq:isi-fp}. In the stationary regime, the firing rate is constant and denoted by $\nu_0$, and the density factorizes as
\begin{equation}
    q^0(v,\tau) = \nu_0\, g^0(v,\tau).
\end{equation}
The function $g^0$ satisfies a time-homogeneous renewal equation in the variables $(v,\tau)$:
\begin{equation}
    \partial_\tau g^0(v,\tau)
    = \mathcal{L}_0 \, g^0(v,\tau)
      + \delta_0(\tau)\,\delta_H(v),
\end{equation}

Taking the Laplace transform of $g^0$ with respect to $\tau$ yields
\begin{align}
    (s - \mathcal{L}_0)\,\tilde{g}^0(v,s)
        &= \delta_H(v), \\
    \Rightarrow\quad 
    \tilde{g}^0(v,s)
        &= G^{\mathrm{A}}(v, H; s),
\end{align}

Evaluating the flux of $\tilde{g}^0$ at threshold provides the stationary ISI distribution:
\begin{equation}
    \flux{\tilde{g}^0}(s)
    = \tilde{\rho}(s).
\end{equation}
Thus, as expected, the flux of $g^0$ through the threshold recovers the Laplace-transformed stationary ISI density.

\subsection{Stationary moment equations}
We now turn to the stationary hierarchy of ISI moments obtained from Eq.~\eqref{eq:moments}.  
Let $T_n^0(v)$ denote the $n$-th stationary marginal moment.  
These satisfy
\begin{align}
    T_0^0(v) &= \phi_0(v), \\
    -\mathcal{L}_0 \, T_n^0(v) &= n\,T_{n-1}^0(v),
\end{align}
Solving this hierarchy using the Green's function $G^{\mathrm{A}}$ yields the integral representation
\begin{align}\label{eq:staz-moment-green}
    T_n^0(v)
        &= \int_{\vmin}^{\theta} G^{\mathrm{A}}(v,y;0)\,T_{n-1}^0(y)\,dy \\
        &= n\,D_0^{-1}\, w(v) \notag
           \int_{\vmin}^{\theta} T_{n-1}^0(y)
           \left[\int_{v\lor y}^{\theta} w^{-1}(z)\,dz\right] dy \\
        &= n\,D_0^{-1}\, \notag
            w(v)\int_{v}^{\theta} w^{-1}(y)
            \left[\int_{\vmin}^{y} T_{n-1}^0(z)\,dz\right] dy,
\end{align}
where $w$ denotes the Wronskian associated with the homogeneous operator $\mathcal{L}_0$. Applying the boundary flux operator $-D_0\,\partial_v|_{v=\theta}$ to Eq.~\eqref{eq:staz-moment-green} yields an explicit formula for the stationary ISI moments:
\begin{align}\label{staz-moment}
    m_n^0
        = \frac{\flux{T_n^0}}{\nu_0}
        = \frac{n}{\nu_0}
          \int_{\vmin}^{\theta} T_{n-1}^0(y)\,dy.
\end{align}
This provides a compact integral characterization of all stationary ISI moments in terms of solutions of the recursive hierarchy for $T_n^0$.

\subsection{Relaxation to stationary state}

\begin{figure*}[!ht]\label{fig:relax}
\centering
\includegraphics[width=175mm]{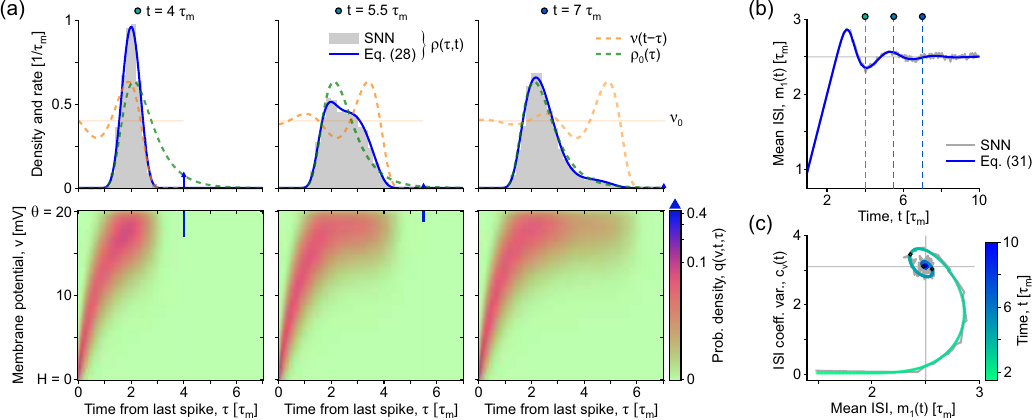}
\caption{
   Relaxation to the stationary state in drift-dominated regime for a network of uncoupled leaky integrate-and-fire neurons (LIF, f(v) = -v): $\mu_0=21 \, \text{mV}$, $D_0=14.15 \, \text{mV}^2$, $H=0 \, \text{mV}$, $\theta=20 \, \text{mV}$.
   (a) Top: time-dependent ISI distribution, comparing \EqRef{eq:relax-H} with spiking neural network simulation (SNN $N=10^4$ neurons). 
   Bottom: population density $q(v,\tau,t)$. 
   (b) Temporal dynamics of mean ISI, comparison between theory and SNN simulation. 
   (c) Temporal trajectory of mean ISI and coefficient of variation. 
   Comparison between theory and SNN simulation.}
\label{fig:1}
\end{figure*}

We analytically solve the relaxation dynamics of an uncoupled neuronal population with an initial state $q(v,\tau,t=0) = \delta_{v_0}(v)\delta_{\tau_0}(v)$, and derive an explicit expression for the time-dependent inter-spike interval (ISI) distribution. 
Here, we focus in particular on the case $(v_0, \tau_0) = (H, 0)$, which corresponds to the relaxation of the population immediately after all neurons are synchronously reset to the reset potential $v=H$---for instance, following an externally triggered burst of activity.
Additional derivations and details are provided in the Appendix. In this case, the time dependent ISI distribution is given by:
\begin{align} \label{eq:relax-H}
\rho(\tau, t) 
   &= \frac{\nu_q(\tau,t)}{\nu(t)}  =  \\ 
   &=\frac{1}{\nu(t)}\left(\rho(\tau)\nu(t-\tau)+ \delta_{t}(\tau)\rho(\tau)\right)  \notag
\end{align}
where $\rho(\tau):= \rho(\tau|H)$ is the stationary state ISI distribution from the renewal theory.
Taking the Laplace transform in $t$ and $\tau$ we obtain:
\begin{align}
    \flux{\widehat{q}}(s, k) = \int_0^\infty\!\!\int_0^\infty \!\!e^{-s\tau - \kappa t}\nu_q(\tau, t) d\tau dt=  \frac{\tilde{\rho}(s+k)}{1-\tilde{\rho}(k)}.
\end{align}
This expression has poles at $\{\lambda_i: \rho(\lambda_i) = 1\}$ and at $\{\mu_i: \rho(\mu_i) = 0\}$, corresponding to the eigenvalues of the Fokker--Planck operator under reinjection and absorbing boundary conditions, respectively \cite{Vinci2024}.
From \eqref{eq:moments} we also obtain a closed-form expression for the relaxation of the moments:
\begin{align}
    \flux{\widehat{T_n}}(k) = (-1)^n\frac{\tilde{\rho}^{(n)}(k)}{1-\tilde{\rho}(k)}.
\end{align}
Evaluating the principal part of this expression at the poles, we can invert the Laplace transform and find a power series expansion for the relaxation of the moment fluxes $\flux{T_n}$. 
Here we report the expression for the first two moments:
\begin{align}
\label{eq1}
    m_1(t) &= \frac{1}{\nu(t)}\sum_{i} e^{\lambda_i t} - e^{\upmu_i t} \to \frac{1}{\nu_0}\\
     m_2(t) &= \frac{2}{\nu(t)}\!\sum_{i} \bigg[\Res_{k=\lambda_i}\frac{\tilde{\rho}(k)}{(k-\lambda_i)^3}{\Res_{k=\lambda_i}\frac{1}{1-\tilde{\rho}(k)}}\bigg]e^{\lambda_i t}\ -\notag \\ &- \frac{2}{\nu(t)}\!\sum_{i}\bigg[t + \!\left(1-\!\Res_{k=\upmu_i}\frac{\tilde{\rho}(k)}{(k-\upmu_i)}\right)\!\!\left(\Res_{k=\upmu_i}\!\tilde{\rho}(k)\right)^{\!\!-1}\!\bigg]e^{\upmu_i t} \notag
\end{align}

\section{Linear Response theory}

Next, we analyse first-order perturbations of the stationary state in the frequency domain. We consider a small time-dependent modulation of the external current $(\mu(t), D(t)) = (\mu_0, D_0) + (\mu_1, D_1)\varepsilon(t)$. We can then write the FP operator $\mathcal{L} = \mathcal{L}_{0} + \mathcal{L}_{1}\varepsilon(t)$ where $\mathcal{L}_{1} = \mu_1\partial_v + D_1\partial^2_v$. 
We start by considering the first order response of the function $g(v, \tau, t) = g^0(v, \tau) + g^1(v, \tau, t) + o(\varepsilon)$. From \eqref{eq:g-fp} we obtain:
\begin{align}\notag
    \partial_t g^1(v, \tau, t) = \mathcal{L}_0  g^1(v, \tau, t) - \partial_\tau g^1(v, \tau, t) +  \varepsilon(t)\mathcal{L}_1 g^0(v, \tau), 
\end{align}
Then, considering the Laplace transform in $t$ and $\tau$:
\begin{align}
    \left(s+\kappa - \mathcal{L}_0\right)\widehat{g}^1(v,s, \kappa) =  \widehat{\varepsilon}(\kappa)\mathcal{L}_1 \tilde{g}^0(v, s)
\end{align}
Which can be solved using the Green function $G^A$:
\begin{align}\notag
    \mathrm{H^{g}}(s, \kappa) = &\frac{\nu_{\widehat{g}^1}(s, \kappa)}{\epsilon(k)} = \int_\vmin^\theta \mathcal{L}_1 \tilde{\rho}(s + \kappa| y) G^\mathrm{A}(y,H;s) dy =\\ = & \int_\vmin^\theta\frac{ \mathcal{L}_1 \psi(y; s+\kappa)}{\psi(\theta; s+\kappa)} G^\mathrm{A}(y,H;s) dy
\end{align}
From this, we can get the response function for the density $\flux{q^1}$: 
\begin{align}\label{lr-q}
    \mathrm{H^{q}}(s, \kappa) = &{\nu_0}\mathrm{H^{g}}(s, \kappa)  + \tilde{\rho}(s+\kappa)\mathrm{H}^{\nu}(\kappa)
\end{align}
Where $\mathrm{H}^{\nu}(\kappa) = \nu^1(\kappa)/\varepsilon(\kappa)$ is the standard linear response function for the population firing rate. Evaluating Equation $\eqref{lr-q}$ at $s=0$ we get an equivalent expression for  $\mathrm{H}^{\nu}(\kappa)$: 
\begin{align}
    \mathrm{H}^{\nu}(\kappa) = \int_\vmin^\theta\frac{ \mathcal{L}_1 \psi(y; \kappa)}{\psi(\theta; \kappa) - \psi(H; \kappa)} \phi_0(y) dy
\end{align}
Finally as $\flux{q}  = \nu_0 + \nu_0 \rho^1 + \nu^1 \rho + o(\varepsilon)$, the linear response of the ISI distribution is given by: 
\begin{align}\label{lr-q}
    \mathrm{H^{\rho}}(s, \kappa) = \mathrm{H^{g}}(s, \kappa) + \frac{1}{\nu_0}\mathrm{H}^{\nu}(\kappa)\left[\tilde{\rho}(s+\kappa) - \tilde{\rho}(s)\right] \, .
\end{align}


\begin{figure}[!ht]
\centering
\includegraphics[width=85mm]{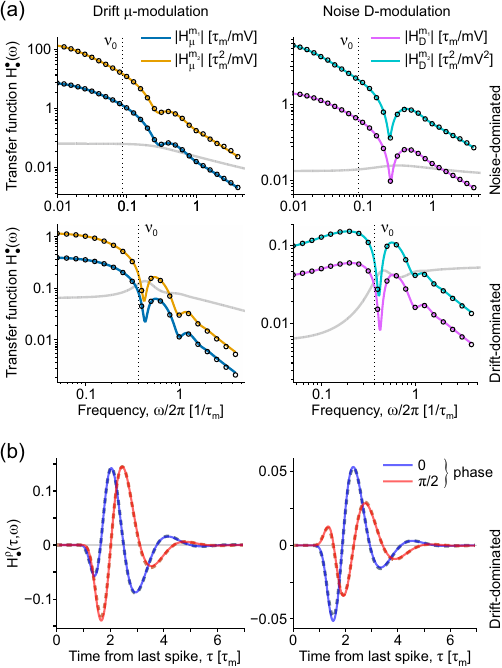}
\caption{
   Linear response and transfer functions of first ISI moments.
   (a) Transfer function for the first two moments. 
   Top: drift-dominated regime (same parameters as \FigRef{fig:relax}) and noise-dominated regime ($\mu = 16.314 \, \text{mV}/\tau_m$, $D_0=14.25 \, \text{mV}^2/\tau_m$, $H=10 \, \text{mV}$, $\theta=20 \, \text{mV}$). comparison between theory (solid lines) and moment hierarchy numerical integration \EqRef{eq:moments}. 
   Solid grey line represents the firing rate transfer function $\mathrm{H}^\nu$. 
   (b) Linear response theory for the ISI distribution in drift-dominated regime, $\omega/2\pi = 0.4\tau_m$. 
   Left: drift modulation, right: noise modulation. 
   Comparison between theory and numerical integration of \EqRef{eq:isi-fp}.}
\label{fig:2}
\end{figure}

\subsection{Linear response from the moment hierarchy}

To derive the linear response of the ISI moments, we expand the moment hierarchy
equations~\eqref{eq:moments} to first order in the input modulation.  
Let 
\[
T_n(v,t)=T_n^0(v)+\varepsilon\,T_n^1(v,t)+o(\varepsilon)
\]
denote the perturbed marginal moments, and let $\widehat{T}_n^1(v,\kappa)$ be their bilateral Laplace transforms in time.  
The corresponding linear response of the ISI moments follows from the identity
\begin{align}
\mathrm{H}^{m_n}(\kappa)
    = (-1)^n\,\partial_s^{\,n}\mathrm{H}^{\rho}(s,\kappa)\big|_{s=0},
\end{align}
which links the moment response function to the ISI response function.

The linearized equations for the perturbations $\widehat{T}_n^1$ are obtained by applying the adjoint Green's function $G^{\mathrm{A}}$ of the stationary operator and collecting all first-order terms.  
For the zero-th element of the hierarchy we obtain
\begin{align}
    \frac{\widehat{T}_0^1(v,\kappa)}{\widehat{\varepsilon}(\kappa)}
        = & \int_{\vmin}^{\theta} 
           G^{\mathrm{A}}(v,y;\kappa)\,
           \mathcal{L}_1\phi_0(y)\,dy \\
          & + G^{\mathrm{A}}(v,H;\kappa)\,\mathrm{H}^{\nu}(\kappa),\notag
\end{align}
For $n\ge 1$ the hierarchy propagates through the recursion
\begin{align}
    \frac{\widehat{T}_n^1(v,\kappa)}{\widehat{\varepsilon}(\kappa)}
        &= \int_{\vmin}^{\theta} 
           G^{\mathrm{A}}(v,y;\kappa)\,
           \mathcal{L}_1 T_n^0(y)\,dy \\
        &\quad + \int_{\vmin}^{\theta} 
           G^{\mathrm{A}}(v,y;\kappa)\,
           \frac{\widehat{T}_{n-1}^1(y,\kappa)}{\widehat{\varepsilon}(\kappa)}\,dy. \notag
\end{align}

Substituting these expressions into the definition of the ISI moments yields the linear 
moment response function
\begin{align}
\mathrm{H}^{m_n}(\kappa)
    &= \frac{1}{\nu_0}
       \int_{\vmin}^{\theta}
       \mathcal{L}_1\tilde{\rho}(\kappa|y)\,T_n^0(y)\,dy \\
    &\quad + \frac{1}{\nu_0}
       \int_{\vmin}^{\theta}
       \tilde{\rho}(\kappa|y)\,
       \frac{\widehat{T}_{n-1}^1(y,\kappa)}{\widehat{\varepsilon}(\kappa)}\,dy
       - \frac{m_n^0}{\nu_0}\,\mathrm{H}^{\nu}(\kappa), \notag
\end{align}
which provides a closed expression for the frequency-domain linear response of the ISI
moments in terms of the stationary hierarchy, the stationary ISI distribution, and the 
perturbation operator $\mathcal{L}_1$.

Figure~\ref{fig:2} illustrates the linear response properties predicted by the theory.  
Panel~(a) shows the transfer functions of the first two ISI moments in both a drift–dominated and a noise–dominated regime.  
In each case, the theoretical prediction (solid lines) closely matches the result obtained by numerically integrating the moment hierarchy~\eqref{eq:moments}. Panel~(b) compares the full ISI response function $\mathrm{H}^\rho$ (solid lines) with numerical integration of the population equation~\eqref{eq:isi-fp} (dashed lines) for a representative modulation frequency ($\omega/2\pi = 0.4/\tau_m$). Both drift and noise modulations are shown.  
In each case, the linear theory accurately reproduces the phase and amplitude of the ISI perturbation, confirming that the response of the full ISI density is correctly captured. 

\section{Nonlinear dynamics}

\begin{figure}[!ht]
\centering
\includegraphics[width=86mm]{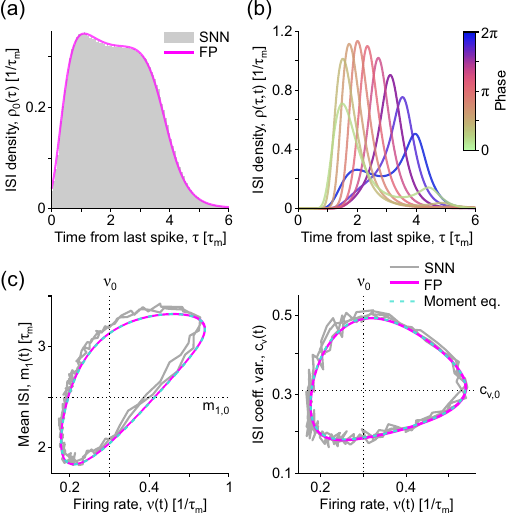}
\caption{
   Synchronous (limit cycle) irregular state in an excitatory population of LIF neurons. 
   (a) ISI distribution $\rho_0(\tau) = \langle \nu_q(\tau,t)\rangle_t/\langle \nu(t)\rangle_t$. 
   Comparison between population dynamics, numerical integration of \EqRef{eq:isi-fp} and SNN simulation ($N=5\cdot10^4$). 
   (b) Phase-dependent ISI distribution. 
   (c) Joint trajectory of firing rate, ISI mean (left) and coefficient of variation (right). 
   Comparison between population dynamics \eqref{eq:isi-fp}, moment hierarchy from \EqRef{eq:moments}) and SNN simulation ($N=5\cdot10^4$). 
   Parameters: $J=0.1 \, \text{mV}$, $K=1000$, $\mu_\text{ext} = 21 \, \text{mV}/\tau_m$, $D_\text{ext}=1.33 \, \text{mV}/\tau_m$, $H=0 \, \text{mV}$, $\theta=20\, \text{mV}$, $\delta = 0.05 \, \tau_m$, $\tau_\delta = 0.1 \tau_m$.}
\label{fig:3}
\end{figure}

Finally, we validate our theoretical results through direct numerical simulations, showing that the proposed population equation accurately captures the time evolution of the ISI distribution even when the activity is far from the stationary state. 
Here, we simulate a coupled population of neurons with deterministic synaptic efficacies $J$ and a distribution of delays in the axonal trasmission of spikes $p_D(x) = \Theta(x > \delta) \exp(-x/\tau_d - \delta)$, where $\Theta$ denotes the Heaviside step function.
Under mean-field approximation this leads the following infinitesimal mean and variance of the synaptic current
\begin{align}
   \mu(t) &= K J (p_D * \nu)(t) + \mu_\text{ext} \\
   D(t)   &= 2 K J^2 (p_D * \nu)(t) + D_\text{ext} \, ,
\end{align}
where $K$ is the number of presynaptic contacts.

We focus on two representative regimes in which the population converges to a stable limit cycle: an excitatory synchronous--irregular oscillation, where all neurons fire on every cycle, and an inhibitory synchronous--regular oscillation, where only a fraction of the neurons fire on each cycle. 
In both cases, we compare the predictions of the population density \EqRef{eq:isi-fp} and the moment hierarchy \eqref{eq:moments} with large-scale spiking-network (SNN) simulations.

Figure~\ref{fig:3} illustrates the excitatory limit cycle.  
Panel~(a) compares the ISI distribution obtained from long-time averages with the prediction of the population equation and with SNN simulations ($N=5\cdot10^4$).  
The ISI distribution reflects a fully synchronous population in which all neurons fire on every cycle but at different phases.  
Panel~(b) shows the phase-resolved ISI distribution, revealing the temporal organisation of spikes within the cycle.  
Panel~(c) tracks the firing rate, mean ISI, and coefficient of variation over a full period.  
Both the population equation and the moment hierarchy match the microscopic dynamics, confirming that the ISI evolution is accurately captured even far from stationarity.

\begin{figure}[!ht]
\centering
\includegraphics[width=83mm]{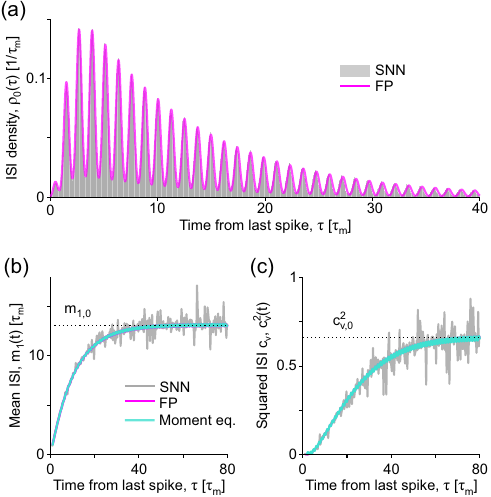}
\caption{
   Synchronous (limit cycle) irregular in an inhibitory population of LIF neurons. 
   (a) ISI distribution $\rho_0(\tau) = \langle \nu_q(\tau,t)\rangle_t/\langle \nu(t)\rangle_t$. 
   Comparison between population dynamics, numerical integration of \EqRef{eq:isi-fp}) and SNN simulation ($N=5\cdot10^4$). 
   (b) Time trajectory of mean ISI. 
   Comparison between population dynamics \eqref{eq:isi-fp}, moment hierarchy \eqref{eq:moments} and SNN simulation ($N=5\cdot10^4$). 
   (c) Same as (b) but tracking squared coefficient of variation. 
   Parameters: $J=-0.1 \, \text{mV}$, $K=1000$, $D_\text{ext} = 8 \, \text{mV}$, $\delta = 0.4 \, \tau_m$, $\tau_\delta = 0\, \tau_m$. 
   Other parameters as in \FigRef{fig:3}. 
}
\label{fig:4}
\end{figure}

The inhibitory limit cycle, shown in Fig.~\ref{fig:4}, displays qualitatively different ISI statistics.  
Because inhibition prevents a subset of neurons from firing on each cycle, the ISI distribution becomes highly structured and multimodal, with peaks corresponding to different numbers of skipped cycles.  
Panel~(a) shows that the population equation reproduces this complex structure with high accuracy.  
Panels~(b) and~(c) track the evolution of the mean ISI and the squared coefficient of variation, again showing excellent agreement between the population equation, the moment hierarchy, and SNN simulations.

Overall, these simulations confirm that the population dynamics equation reliably describes ISI statistics in regimes where renewal-based approximations break down and the population is far from the stationary state.  
Moreover, the moment hierarchy offers an efficient complementary description of ISI statistics, accurately capturing moment dynamics at a computational cost comparable to integrating the one-dimensional Fokker--Planck equation.

\section{Discussion}




In summary, we developed an analytic framework for characterizing the time evolution of inter-spike interval (ISI) distributions in non-stationary neuronal populations. Starting from the population density equation, we formulated a two-dimensional Fokker--Planck description that jointly tracks membrane voltage and time since the last spike, providing direct access to the ISI distribution under arbitrarily time-dependent inputs.

Numerical integration of this equation shows excellent agreement with large-scale spiking-network simulations, including regimes that strongly depart from stationarity, such as recurrently generated limit cycles. This demonstrates that the framework captures ISI dynamics beyond the reach of classical renewal-based approaches.

Building on this formulation, we derived a one-dimensional hierarchy governing the moments of the ISI distribution and performed a first-order perturbative expansion around the stationary state. This yields a compact analytic expression for the linear response of ISI statistics to weak temporal modulation, offering a reduced yet accurate description of ISI dynamics.

More broadly, the methodology provides a tractable route to analyzing threshold-crossing statistics in driven excitable systems, extending its relevance beyond neuronal models to a wider class of first-passage problems in statistical physics.

\begin{acknowledgments}
This research has received funding from the Italian National Recovery and Resilience Plan (PNRR), M4C2, funded by the European Union - NextGenerationEU (Project IR0000011, CUP B51E22000150006, `EBRAINS-Italy'), to MM.
\end{acknowledgments}

\appendix

\section{Details on the Green function}
The Green function $G^{\mathrm{A}}(x,y; s)$ for the Fokker-Planck operator $\mathcal{L}_0$ with absorbing boundary conditions can be expressed as a function of the solutions of the homogeneous equations: 
\begin{align} \notag
     \mathcal{L}_0 f_i(\cdot; s) = sf_i(\cdot; s)\qquad 
     \mathcal{L}_0^+ \psi_i(\cdot; s) = s\psi_i(\cdot; s) \quad i\in\{1,2\}.
\end{align}
Where $\mathcal{L}_0^+$ is the adjoint operator. such that: 
\begin{align}
     \partial_v\psi_i(v; s)|_{v=\alpha} = \delta_{i1}\qquad 
     \psi_i(\alpha; s) = \delta_{i2} \quad i\in\{1,2\}
\end{align}
\begin{widetext}
Then the Green's function has the form: 
\begin{align}
    G^{\mathrm{A}}(x,y; s) = 
    \begin{cases}
        f_1(x; s)\cdot a_-(y,s) + f_2(x; s)\cdot b_-(y,s) & x<y \vspace{1em}\\ 
        f_1(x; s)\cdot a_+(y,s) + f_2(x; s)\cdot b_+(y,s) & x>y 
    \end{cases}
\end{align}
The coefficients $a_\pm(y,s), b_\pm(y,s) \in \mathbb{C}$, can be found by imposing the conditions:
\begin{align}
    \partial_xG^{\mathrm{A}}(x,y; s)|_{x=\alpha} = 0, \quad
    G^{\mathrm{A}}(\theta,y; s) = 0,\quad
    \mathcal{L}_0 G^{\mathrm{A}}(\cdot,y; s) = \delta_y.
\end{align}
Solving the associated linear system, we find: 
\begin{align*}
    G^{\mathrm{A}}(x,y; s) = \frac{1}{D_0}
    \begin{cases}
        f_1(x; s)\cdot \frac{\psi_1(\theta;s)\psi_2(y;s) -\psi_2(\theta;s)\psi_1(y;s)}{\psi_1(\theta;s)} & x<y \vspace{1em}\\ 
        f_2(x;s)\cdot \frac{\psi_1(y;s)\psi_1(\theta;s)}{\psi_1(\theta;s)} - f_1(x;s)\cdot \frac{\psi_2(\theta;s)\psi_1(y;s)}{\psi_1(\theta;s)} & x>y 
    \end{cases}
\end{align*}

\end{widetext}

\section{Relationship with Siegert equation}

Historically, the inter-spike interval (ISI) distribution has been studied using a “forward-in-time” approach: for a neuron starting at a given state $v$ one computes the distribution of the first passage time to a threshold 
$\theta$, typically using the backward Kolmogorov operator $\mathcal{L}^+$. In his seminal work, Siegert used this framework to derive a recursive system of equations for the moments of the stationary ISI distribution:
\begin{align}\label{siegert-moments}
    -\mathcal{L}^+_0{m^0_n(y)} = n\,m_{n-1}^0(y)
\end{align}
\begin{align}
     m_n^0(y) = nD_0^{-1}\int_y^\theta w^{-1}(z) \int_\alpha^z w(v)m_{n-1}^0(v)dvdz.
\end{align}
\begin{align}
     w(v) = f_1(v)\partial_v f_2(v) - f_2(v) \partial_v f_1(v)
\end{align}
in a non-stationary setting, the “forward-in-time” approach requires knowledge of the future input current, making it difficult to extend to systems—such as coupled neural networks—where the input must be determined self-consistently from the population activity, i.e., from the evolving probability flux.

In contrast, we adopt a “backwards-in-time” perspective, describing the inter-spike interval distribution of the neurons in the population that are currently firing. 
Here we show that in the case of stationary currents, the two formal descriptions are equivalent. 
While the function $m^0_n(v)$ describes n-th moment of the distribution of waiting time until the next spike for a neuron with membrane potential $v$, instead the function $T^0_n(v)$ is related to the n-th moment of the time from the last spike.  
The two variables have the following relationship:
\begin{align}
     \frac{1}{\nu_0}\int_\vmin^\theta T_{n-1}^0(v)dv = \frac{m_n^0}{n} = \int_\vmin^\theta m^0_{n-1}(v)\phi_0(v)dv.
\end{align}

In fact, using Equation \eqref{siegert-moments} and integrating by parts we have: 
\begin{align*}
    n\int_\vmin^\theta m^0_{n-1}(v)\phi_0(v)dv = -\int_\vmin^\theta \mathcal{L}^+_0 m_{n}^0(v)\phi_0(v)dv =\\= \nu_0m_n^0(H) = \nu_0m_n^0 = \int_{\vmin}^\theta T_{n-1}^0(v)dv
\end{align*}
where the last equality follows from Equation \eqref{staz-moment}. Similarly, with the same strategy, we can prove the following relation: 
\begin{align*}
    \int_\vmin^\theta q^0(\tau, v) dv = \int_\vmin^\theta \rho(\tau|v) \phi_0(\tau, v) dv = \nu_0 \mathrm{L}^{-1}\bigg\{\frac{1-\rho(s)}{s}\bigg\}
\end{align*}
where $L^{-1}\{\cdot\}$ is the inverse Laplace transform
\section{Additional details on the age-structured population dynamics}
We decompose the full density $q(v,\tau, t)$ by normalizing the reinjection flux at$ (v, \tau)=(0, H)$ and  write: 
\begin{align}\label{eq:dyn-decomp}
    q(v, \tau, t) = 
        g(v,\tau, t)\nu(t-\tau) + 
        \ell(v,\tau, t)
\end{align}
With initial conditions
\begin{equation}
g(v,\tau,0)=0,
\qquad
\ell(v,\tau,0)=q(v,\tau,0),
\end{equation}
where $\ell$ accounts for neurons that have not fired during the interval 
$[0,t]$, with support restricted to $\tau \ge t$. At the same time, $g$ describes the probability density of a neuron that last fired at time $t - \tau > 0$ and is thus defined only for $\tau < t$. We may then write 
$$q(v, \tau, t) = \nu(t-\tau)g(v,\tau,t), \forall \tau < t$$
Substituting this decomposition into Eq.~\eqref{eq:isi-fp} yields:
\begin{align}
    &\nu(t-\tau)\partial_tg(v, \tau, t) + g(v, \tau, t)\partial_t \nu(t-\tau) = \notag\\  &\nu(t-\tau)\mathcal{L} \, g(v, \tau, t) - g(v,\tau,t)\partial_\tau \nu(t-\tau)-\\ - &\nu(t-\tau)\partial_\tau g(v,\tau, t) + \nu(t-\tau)\delta_H(v)\delta_0(\tau) \notag\quad \forall \tau < t
\end{align}
Using the identity $\partial_t \nu(t-\tau) = - \partial_\tau$, this expression simplifies to Eq. ~\eqref{eq:g-fp}. 
\subsection{Relaxation dynamics}

For stationary diffusion coefficients, the population dynamics in Eq.~\eqref{eq:isi-fp} can be solved explicitly.  
We first consider the elementary initial condition 
\begin{align}
    q(v,\tau,0) = \ell(v,\tau,0) = \delta_y(v)\,\delta_u(\tau).
\end{align}

Using the decomposition~\eqref{eq:dyn-decomp}, the two contributions read
\begin{align}
    \nu_\ell(\tau,t) &= \delta_{u+t}(\tau)\,\rho(t|y), \\
    \nu_g(\tau,t) &= \rho(\tau), \qquad \tau < t,
\end{align}
from which the relaxation ISI distribution follows as
\begin{align}
    \rho(\tau,t)
    = \frac{1}{\nu(t)}
      \begin{cases}
        \rho(\tau)\,\nu(t-\tau), & \tau < t, \\[4pt]
        \delta_{u+t}(\tau)\,\rho(t|y), & \tau \ge t.
      \end{cases}
\end{align}

For a generic initial condition $q(v,\tau,0)=q_*(v,\tau)$, the same construction yields
\begin{align}
    \rho(\tau,t)
    = \frac{1}{\nu(t)}
      \begin{cases}
        \rho(\tau)\,\nu(t-\tau), & \tau < t, \\[6pt]
        \displaystyle \int_{\vmin}^{\theta} \rho(t|y)\,q_*(y,t-\tau)\,dy, 
        & \tau \ge t.
      \end{cases}
\end{align}

\subsection{Refractory density and hazard rate}

We can rewrite equation \eqref{eq:refractory} as: 
\begin{align}
\partial_t r(\tau, t) &= - \partial_\tau r(\tau, t) - h(\tau, t)r(\tau, t) + \nu(t)\delta_{0}(\tau), \\
h(\tau, t) &:= \frac{\nu_g(\tau, t)}{I(\tau, t)},
\end{align}
where $h(\tau, t)$ is the time-dependent hazard rate, representing the instantaneous firing probability of neurons that last emitted a spike 
$\tau$ units of time ago. The quantity
\begin{align}
I(\tau, t) = \int_{\vmin}^\theta g(v, \tau, t) dv
\end{align}
defines the survival function, which denotes the fraction of neurons that fired exactly $\tau$ units of time ago and have not fired again since then. Its temporal evolution is governed by: 
\begin{align}
\partial_t I(\tau, t) = -\partial_\tau I(\tau, t) - \nu_g(\tau, t) + \delta_0(\tau).
\end{align}

\bibliographystyle{apsrev4-2}
\bibliography{RefLibrary}

\end{document}